\documentclass{aa}  
\usepackage{graphicx}
\usepackage{txfonts}
\usepackage{natbib}
\bibpunct{(}{)}{;}{a}{}{,} 

\begin{document}
   \title{Analysis of 70 Ophiuchi AB including seismic constraints}

   \author{P. Eggenberger
          \inst{1}
	  \and
	  A. Miglio\inst{1}	
          \and
	  F. Carrier\inst{2}
 	  \and         
	  J. Fernandes\inst{3}
	  \and		
	  N.C. Santos\inst{4}	 
          }

   \offprints{P. Eggenberger}

   \institute{Institut d'Astrophysique et de G\'eophysique de l'Universit\'e de Li\`ege, 17 All\'ee du 6 Ao\^ut, 	       B-4000 Li\`ege, Belgium\\
              \email{eggenberger@astro.ulg.ac.be; miglio@astro.ulg.ac.be}
\and
             Institute of Astronomy, University of Leuven, Celestijnenlaan 200 B, B-3001 Leuven, Belgium\\
             \email{fabien@ster.kuleuven.be}
         \and
             Observat\'orio Astron\'omico da Universidade de Coimbra e Departamento de Matem\'atica, FCTUC, 		     Portugal \\
             \email{jmfernan@mat.uc.pt}         
         \and
             Centro de Astrof\'isica, Universidade do Porto, Rua das Estrelas, P--4150-762 Porto, Portugal\\
             \email{nuno@astro.up.pt}
             }

   \date{Received; accepted}

 
  \abstract
   {The analysis of solar-like oscillations for stars belonging to a binary system
    provides a unique opportunity to probe the internal stellar structure and to test
    our knowledge of stellar physics.	
    Such oscillations have been recently observed and characterized for the A component  	   of the 70~Ophiuchi system.}
   {We determined the global parameters of 70~Ophiuchi AB using the new 
asteroseismic measurements now available for \object{70~Oph~A} and tested the input physics introduced in stellar evolution codes.}
   {Three different stellar evolution codes and two different calibration methods were
    used to perform a comprehensive analysis of the 70~Ophiuchi system.}
   {A model of 70~Ophiuchi AB that correctly reproduces all observational constraints available
for both stars is determined. An age of $6.2 \pm 1.0$\,Gyr is found with an initial
helium mass fraction $Y_{\mathrm{i}}=0.266 \pm 0.015$ and an initial metallicity 
$(Z/X)_{\mathrm{i}}=0.0300 \pm 0.0025$ when atomic diffusion is included and
a solar value of the mixing-length parameter assumed. A precise and independent determination of the
value of the mixing-length parameter needed to model \object{70~Oph~A} requires accurate
measurement of the mean small separation, which is not available yet. Current asteroseismic observations, however, suggest 
that the value of the mixing-length parameter of \object{70~Oph~A} is lower or equal to the solar calibrated value. 
The effects of atomic diffusion and of the choice of the adopted solar mixture were
also studied. We finally found that the different evolution codes and calibration methods
we used led to perfectly coherent results.}
   {}


   \maketitle
%

\section{Introduction}

The solar five-minute oscillations have provided a wealth of information
on the internal structure of the Sun. 
These results stimulated various attempts to detect
a similar signal on other solar-type stars by photometric or equivalent
width measurements. In past years, the stabilized spectrographs developed for extra-solar planet search have finally achieved the accuracy needed to detect solar-like oscillations on other stars
\citep[see e.g.][]{bed07}. 
A major difficulty is the confrontation between these observations
and theoretical models. The classical observational measurements available 
for an isolated star (effective temperature, metallicity, luminosity) combined 
with the oscillation frequencies provide strong constraints to the global parameters of the star but are often not sufficient to unambiguously determine a unique model and to really test the input physics included in the computation of the stellar models 
\citep[see for example the case of the isolated star \object{$\beta$~Virginis},][]{egg06}.
In this regard, binary stars constitute ideal asteroseismic targets in order 
to test our knowledge of stellar physics. 
In the case of a binary system, we can indeed assume that both stars share the same age and initial chemical composition.
The additional constraints imposed by this assumption are then extremely valuable to accurately determine the
global properties of the stars. Moreover, in the case of a binary system, 
the masses of both components are accurately known by combining visual and spectroscopic orbits.

Unfortunately, individual frequencies of solar-like oscillations have only been identified for stars belonging to two different binary systems: the G2 dwarf star  \object{$\alpha$~Cen A} \citep{bou02,bed04,baz07}, the K1 dwarf star
 \object{$\alpha$~Cen B} \citep{car03,kje05} and the F5 IV-V star  \object{Procyon A} \citep[][]{mar04,egg04_pro,mos07}.
Recently, \cite{car06} reported the detection of solar-like oscillations on the bright K0 dwarf star \object{70~Ophiuchi A} (\object{HD~165341A}) belonging to the  
nearby visual binary system 70~Ophiuchi, which is among the first discovered binary systems.
The aim of the present paper is to perform a comprehensive calibration of the 70~Ophiuchi 
system that includes the new seismological data available for the A component.
In addition to the determination of the 70~Ophiuchi~AB global parameters, we also
test and compare the theoretical
tools used for the modeling of stars for which p-modes frequencies are detected
by performing this analysis with three different stellar evolution codes and two different calibration methods.
This work therefore takes place in the continuity of the work undertaken by the Evolution
and Seismic Tools Activity (ESTA) group within the CoRoT mission \citep{mon06}.

The observational constraints available for the 70~Ophiuchi system are summarized in Sect.~2.
The input physics and the computational methods used for the calibrations are described in Sect.~3. The results are presented in Sect.~4, while the conclusion is given in Sect.~5.

\section{Observational constraints}
\label{obs}

\subsection{Binarity}

\begin{figure}[htb!]
 \resizebox{\hsize}{!}{\includegraphics{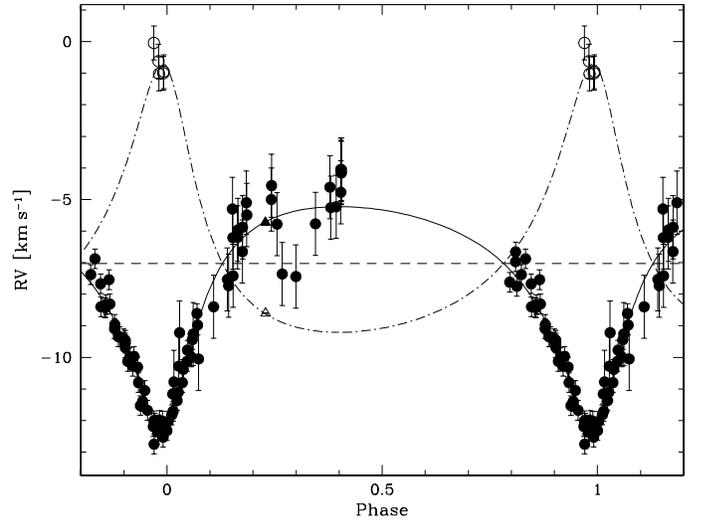}}
  \caption{Radial velocity orbit of the 70~Ophiuchi system. Full and open symbols correspond to
  \object{70~Oph~A} and B, respectively. Triangles indicate the new \textsc{Harps} measurements, while
  circles indicate previous measurements.}
  \label{ophu_vr}
\end{figure}

\begin{figure}[htb!]
 \resizebox{\hsize}{!}{\includegraphics[angle=-90]{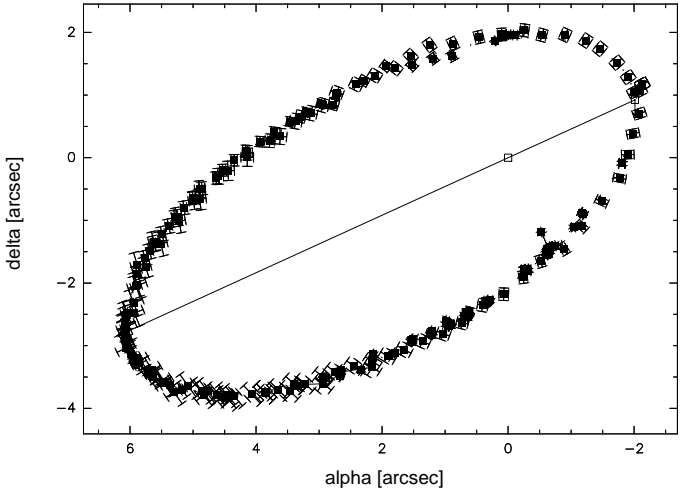}}
  \caption{ Apparent visual orbit of the 70~Ophiuchi system. The solid line indicates the line of nodes, while the square indicates the position of the periastron.}
  \label{ophu_vis}
\end{figure}

70 Ophiuchi is a very well studied stellar system composed of a K0 and K5 dwarf star with an orbital period of
88.38 yr \citep{pou00}. It is known as a binary by both spectroscopy
and speckle interferometry. Radial-velocity curves were first obtained by \cite{bat84} and
\cite{bat91}, while the whole orbit (speckle and radial velocity) was published by \cite{pou00}. 
This binary system was observed over 6 nights in July 2004 with the \textsc{Harps} spectrograph mounted
on the 3.6-m telescope at La Silla Observatory (ESO, Chili) \citep{car06}.
Combining our new radial-velocity measurements
with previous ones \citep{pou04}, we use the ORBIT code made available by T.
Forveille and developed in Grenoble \citep{for99} to simultaneously fit the radial
velocities and the speckle measurements (see Fig.~\ref{ophu_vr} and \ref{ophu_vis}). 
Note that the very high precision of the new \textsc{Harps} data is quite important in order to derive an accurate value for $K_2$ and hence a reliable mass for the B component. Moreover, these new data
add an important observation of the radial velocity of \object{70~Oph~B} since only few 
measurements were previously obtained for this component (see Fig.~\ref{ophu_vr}). The values of the derived orbital parameters, of the parallax, and of the masses of both components are listed in Table~\ref{orb}. We note that the value of the parallax derived in this
way ($194.2 \pm 1.2$\,mas) is in good agreement with the value of $195.7 \pm 0.9$\,mas determined
by \cite{sod99} using the Hipparcos astrometry with ground-based observations available at that time.

\begin{table}
\caption{Orbital elements derived for the 70~Ophiuchi system when all available measurements
(speckle and radial velocity) are included (see text for more details).}
\begin{center}
\begin{tabular}{lc}
\hline
\hline
Parameter  &  Value \\ \hline
$P$ (days)    & 32279 $\pm$ 15\\ 
$T_{0}$ (HJD)& 45809 $\pm$ 14 \\ 
$e$            & 0.5005 $\pm$ 0.0006\\ 
$V_0$ (km s$^{-1}$)& -7.026 $\pm$ 0.015 \\ 
$i$            & 121.1 $\pm$ 0.1\\ 
$\omega$ (deg)& 193.4 $\pm$ 0.3\\ 
$\Omega$ (deg)& 121.7 $\pm$ 0.2\\ 
$K_1$ (km s$^{-1}$) & 3.51 $\pm$ 0.04\\
$K_2$ (km s$^{-1}$) & 4.25 $\pm$ 0.05\\
$a$ (arcsec)  &4.526 $\pm$ 0.007\\ 
$M_1$ (M$_{\odot}$)        & 0.89 $\pm$ 0.02\\ 
$M_2$ (M$_{\odot}$)        & 0.73 $\pm$ 0.01\\ 
$\pi$ (mas) & 194.2 $\pm$ 1.2 \\
\hline
\label{orb}
\end{tabular}
\end{center}
\end{table}

\subsection{Effective temperatures and chemical composition}
\label{tchim}

The effective temperature and iron abundance of the two stars were
obtained from our own \textsc{Harps} spectra
by following the methodology and line-lists used
in \cite{san04_ab}. This determination makes use of
an iteractive procedure based on iron excitation and ionization
equilibrium. We refer to this paper for more details.
The results of this method were shown to give excellent
results for solar-type stars, and the derived stellar parameters, namely
the effective temperature, are compatible with those derived
from other methods \citep[e.g. the IR-flux method,][]{cas06}.
For the metallicity of the system a mean value of $[$Fe/H$]=0.04 \pm 0.05$ was thus determined
and used for the present calibration. No $\alpha$-element enhancement is suggested
for the 70~Ophiuchi system by current spectroscopic data.
 Concerning the effective temperature,
we derived a value of $5300 \pm 50$\,K for the A component.
For the B component, an effective temperature of 4205\,K is found with an
uncertainty estimated to 100\,K. 
We however note that our spectroscopic determination is less adapted 
for this kind of cold star.
We thus decided to adopt as $T_{\rm eff}^{\rm B}$ a mean value between our determined spectroscopic temperature of 4205\,K, our photometric deduced value of 4220\,K (see Section~2.3)
and the effective temperature of 4740\,K determined by \cite{luc05}. 
In this way a mean temperature of $T_{\rm eff}^{\rm B} = 4390\,K$ was determined for 
the B component. We also decided to adopt a large error of 200\,K on the effective temperature
of  \object{70~Oph~B}, so that our calibration of the system is only weakly constrained
by this value, which is not precisely determined.
 
\subsection{Luminosities}
\label{lum}

From 1977 to 2005, the 70 Ophiuchi system has been measured in the \textsc{Geneva} photometric system \citep{gol80}
with the photoelectric photometer P7 \citep{bur79} installed first on the Swiss telescope in La Silla (ESO, Chile)
and then on the Belgian telescope at La Palma. Nine measurements were obtained. However, due to the small separation
between components and to too large a diaphragm, these photometric measurements only represent the mean flux
of both components $V_{\rm tot}\,=\,4.012$\,$\pm$\,0.013, $V1_{\rm tot}\,=\,4.779$\,$\pm$\,0.013 and $B2_{\rm tot}\,=\,5.323$\,$\pm$\,0.015\,mag. 
In order to obtain individual photometric measurements of both components, the 70~Ophiuchi system was observed
with the CCD camera C2 installed on the Swiss telescope at La Silla \citep{che06}
between 2005 August 20-23 using very
short exposure times of 0.02, 0.03 and 0.2\,s in the $V$, $V1$ and $B2$ bands, respectively.
A large number of exposures were obtained in order to reduce the noise due in great part to the scintillation:
80, 18 and 12 exposures were thus taken in the $V$, $V1$ and $B2$ filters, respectively. These CCD frames allowed us
to determine the luminosity shifts between both components; we thus found a difference of 1.870\,$\pm$\,0.014, 1.920\,$\pm$\,0.014
and 2.255\,$\pm$\,0.028\,mag in the $V$, $V1$ and $B2$ bands, respectively. Combining these values with the previous ones,
we finally derived mean visual magnitudes $V_{\rm A}$\,=\,4.191\,$\pm$\,0.014 and $V_{\rm B}$\,=\,6.061\,$\pm$\,0.021\,mag
and photometric indexes $(B2-V1)_{\rm A}$\,=\,0.501\,$\pm$\,0.020 and $(B2-V1)_{\rm B}$\,=\,0.836\,$\pm$\,0.040 for the 70 Ophiuchi system.
Note that these indexes lead to an effective temperature of 4220\,K for the B component \citep{ram05}.

Combining the mean magnitudes with the parallax deduced from the radial-velocity and speckle orbits ($\pi$\,=\,194.2\,$\pm$\,1.2), and by using the solar absolute bolometric magnitude $M_{\rm bol, \odot}$\,=\,4.746 \citep{lej98} with the bolometric corrections from \cite{flo96} 
($BC_{\rm A}$\,=\,-0.20\,$\pm$\,0.02 and $BC_{\rm B}$\,=\,-0.69\,$\pm$\,0.17\,mag), the luminosities
are finaly determined with a value of 0.53\,$\pm$\,0.02\,$L_{\odot}$ 
for \object{70~Oph~A} and 0.15\,$\pm$\,0.02\,$L_{\odot}$ for \object{70~Oph~B}.

\begin{table}
\caption{Observational constraints for \object{70~Oph~A} and B.}
\begin{center}
\begin{tabular}{ccc}
\hline
\hline
 & \multicolumn{1}{c}{\object{70~Oph~A}} & \multicolumn{1}{c}{\object{70~Oph~B}} \\ \hline

$\pi$ (mas)& \multicolumn{2}{c}{$194.2 \pm 1.2$}  \\
$M/M_{\odot}$  & $0.89 \pm 0.02$ & $0.73 \pm 0.01$   \\
$V$ (mag) & $4.191 \pm 0.014$  & $6.061 \pm 0.021$  \\
$L/L_{\odot}$ & $ 0.53 \pm 0.02$ & $0.15 \pm 0.02$   \\
$T_{\mathrm{eff}}$ (K)& $5300 \pm 50$ & $4390 \pm 200$  \\
$[$Fe/H$]$ &  \multicolumn{2}{c}{$0.04 \pm 0.05$}  \\
$\Delta \nu_{0}$ ($\mu$Hz) & $161.8 \pm 0.8$  & -  \\
\hline
\label{tab:constraints}
\end{tabular}
\end{center}
\end{table}

\subsection{Asteroseismic constraints}
\label{asc}

Solar-like oscillations in 70~Ophiuchi~A were recently observed by 
\cite{car06} with the \textsc{Harps} echelle spectrograph.
Fourteen oscillation frequencies were detected in the power spectrum between 3 and 6\,mHz 
with amplitudes in the range 11 to 14\,cm\,s$^{-1}$.
Unfortunately, the mono-site nature of the observations, coupled to the low resolution of the time series and to the faint signal-to-noise, do not allow a unique identification of
the detected frequencies. Indeed, the small separation between $\ell$\,=\,2 and 
$\ell$\,=\,0 modes is not clearly detected and the $\ell$\,=\,2 modes cannot 
therefore be unambiguously disentangled from the $\ell$\,=\,0 modes.
Consequently $\ell$\,=\,0,2 can be replaced by $\ell$\,=\,1 modes
and $\ell$\,=\,1 by $\ell$\,=\,2 modes. Although these measurements cannot allow
an unambiguous identification of individual detected oscillation frequencies,   
they provide us with a precise determination of the mean large separation of \object{70~Oph~A}.
By averaging the 4 different large separations observed for the modes
listed as $\ell$\,=\,0 modes in Table~3 of \cite{car06} with an uncertainty
on individual frequencies estimated to half the time resolution, we obtain a mean value
$\Delta \nu_0 = 161.8 \pm 0.8$\,$\mu$Hz that is used in the present computations. 
We note that the error on this value is slightly larger than the uncertainty of
0.3\,$\mu$Hz quoted by \cite{car06} who included large separation averages  
taken between non-consecutive modes.
 
Although the mean small separation of \object{70~Oph~A} cannot be clearly determined from current
asteroseismic measurements, these observations nevertheless suggest that this value
must be included in a frequency interval centered
around the daily alias of 11.57\,$\mu$Hz introduced by the mono-site observations 
and delimited by the frequency resolution of the time-series (2.2\,$\mu$Hz).
Indeed, if the value of the small separation of \object{70~Oph~A} differs from $11.57\pm2.2$\,$\mu$Hz,
we can then reasonably think that it should have been revealed by existing 
asteroseismic measurements. The inclusion of this additional observational constraint in
the calibration of the 70~Ophiuchi system is discussed in detail in Sect.~\ref{freemlt}.

\section{Stellar models}

\subsection{Input Physics}
\label{phys}

Three different stellar evolution codes are used for these computations: 
the Geneva code \citep{mey00,egg07}, the CLES code 
\citep[Code Li\'egeois d'Evolution Stellaire,][]{scu07} and the CESAM code 
\citep[Code d'Evolution Stellaire Adaptatif et Modulaire,][]{mor97}. 

These three stellar evolution codes are described in detail in the above mentioned references.
We simply note that the Geneva code uses the MHD equation of state \citep{hum88,mih88,dap88} 
and the NACRE nuclear reaction rates \citep{ang99}. The CLES code uses the OPAL equation of state
\citep{rog02} in its 2005 version and the NACRE nuclear reaction
rates, while the version of the CESAM code used for the present computations includes
the EFF equation of state \citep{egg73} 
and the nuclear reaction rates given by \cite{cau88}.
All codes use the opacity tables from the OPAL group \citep{igl96} 
complemented at low temperatures with \cite{ale94} opacities
and the standard mixing-length formalism for convection.
The diffusion due to the concentration and thermal
gradients is also included in the computations, 
but the radiative acceleration is neglected as it is negligible for the
structure of low-mass stellar models with extended convective envelopes \citep{tur98}.

A solar calibration has first been performed
with the three evolution codes. A mixing-length parameter $\alpha=1.7998$ and an initial
helium mass fraction $Y_{\rm i}=0.2735$ is then determined with the physics included in
the Geneva code. The input physics used by the CLES code leads to  
$\alpha=1.8544$ and $Y_{\rm i}=0.2755$, while values of $\alpha=1.7985$ 
and $Y_{\rm i}=0.2848$ are found with the CESAM code. We note that the higher value of the
solar calibrated mixing-length parameter found for the CLES code mainly results from the use of \cite{kur98} atmosphere models in the CLES code instead of the grey atmosphere approximation included in the other evolution codes. Likewise, the CESAM code gives a higher value of the
solar calibrated initial helium mass faction, which is mainly due to the use of the
EFF equation of state instead of the OPAL and MHD equations of state introduced in the
CLES and Geneva codes.

\subsection{Calibration methods}

Basically, the calibration of a binary system consists in finding the set of stellar modeling parameters that best reproduces all observational 
data available for both stars.
The characteristics of a stellar model depend on five modeling parameters: 
the mass $M$ of the star, its age (hereafter $t$), 
the mixing-length parameter $\alpha \equiv l/H_{\mathrm{p}}$ for convection and two parameters describing the initial chemical composition of the star. For these two parameters,
we choose the initial helium abundance $Y_{\mathrm{i}}$ and the initial ratio between the mass fraction of heavy elements and hydrogen $(Z/X)_{\mathrm{i}}$. 
This ratio is directly related to the abundance ratio [Fe/H] 
assuming that $\log(Z/X) \cong \mathrm{[Fe/H]} + \log(Z/X)_{\odot}$. In this study,
stellar models are computed with the solar value $(Z/X)_{\odot}=0.0245$ 
given by \cite{gre93} as well as with the new solar mixture of \cite{asp05} 
complemented with the neon abundance of \cite{cuh06}, which results in a lower value $(Z/X)_{\odot}=0.0178$.

The binary nature of the system leads to a precise determination of the mass of both
components. By assuming that both components of a binary system share the same age and
initial chemical composition, three additional constraints are obtained: $t_{\mathrm{A}}=t_{\mathrm{B}}$, $Y_{\mathrm{i}}^{\mathrm{A}}=Y_{\mathrm{i}}^{\mathrm{B}}$ and $(Z/X)_{\mathrm{i}}^{\mathrm{A}}=(Z/X)_{\mathrm{i}}^{\mathrm{B}}$. Consequently, we have to determine a set of seven modeling parameters 
($t$, $M_{\rm A}$, $M_{\rm B}$, $\alpha_{\mathrm{A}}$, $\alpha_{\mathrm{B}}$, 
$Y_{\mathrm{i}}$ and $(Z/X)_{\mathrm{i}}$). In the case of 70 Ophiuchi,
we also assume that the value of the mixing-length parameter is identical for both
components of the system ($\alpha_{\rm A}=\alpha_{\rm B}$). 
This assumption results from the limited number of
observational constraints available for the B component of the system.
Indeed, no asteroseismic measurements are available for \object{70~Oph~B} and its
effective temperature is only determined with a large error of 200\,K (see Table \ref{tab:constraints}). 

Following the work done for the analysis of the $\alpha$~Centauri system,
the determination of the set of 
modeling parameters ($t$, $M_{\rm A}$, $M_{\rm B}$, $\alpha$, 
$Y_{\mathrm{i}}$ and $(Z/X)_{\mathrm{i}}$) leading to the best agreement with the observational constraints is made by using two different calibration methods. 
The first one is based on the computation of a dense grid of stellar models 
as explained in \cite{egg04_acen}, while the
second one uses the gradient-expansion algorithm known as Levenberg-Marquardt method
\citep[see][]{mig05}. 
The reader is refered to these two references 
for a detailed description of the methods.   
In the present study, the grid of stellar models was computed by
varying the masses of 70~Ophiuchi A and B within their observational errors by step of
0.01\,M$_{\odot}$. Concerning the chemical composition, the initial helium abundance 
$Y_{\mathrm{i}}$ was changed between 0.240 and 0.290 by step of 0.002. Models
of \object{70~Oph~B} computed with an initial helium mass fraction larger than 0.290 are indeed found to
be too luminous to correctly reproduce the observed luminosity. 
For the initial value of the ratio between the mass fraction of heavy elements
 and hydrogen $(Z/X)_{\mathrm{i}}$, we first note that the surface metallicities 
 [Fe/H]$_{\mathrm{s}}$ are almost identical for models of 70~Ophiuchi with the same initial composition and different mixing-length parameters $\alpha$.
Moreover, the [Fe/H]$_{\mathrm{s}}$ of the models are mainly sensitive to $(Z/X)_{\mathrm{i}}$ and less to $Y_{\mathrm{i}}$. 
As a result,
the values of $(Z/X)_{\mathrm{i}}$ are directly constrained by the observed surface metallicities; we found that the models matching the observed
metallicities have $(Z/X)_{\mathrm{i}}$ ranging from about 0.027 to 0.033 (i.e. an initial metallicity [Fe/H]$_{\mathrm{i}}$ between 0.04 and 0.13). §
The value of the mixing-length parameter was changed between 0.8 and 1.3$\alpha_{\odot}$
with a typical step of 0.02$\alpha_{\odot}$.
Finally, we note that the age of the models was limited to 15\,Gyr and 
that a model of \object{70~Oph~A} is considered to have the same age as a model
of \object{70~Oph~B} when the difference is lower than $0.01$\,Gyr.

The theoretical low-$\ell$ p-mode frequencies of \object{70~Oph~A} are calculated using the Aarhus adiabatic pulsation package \citep{chr97} for models computed with the Geneva and the CESAM
code, while the Li\`ege oscillation code \citep{scu07_losc} is used for models computed with the CLES code.
The value of the mean large separation is determined for each model of \object{70~Oph~A} 
by making the average of the theoretical separations corresponding to the
observed frequencies (see Sect.~\ref{asc}). 
A $\chi^2$ minimization is then performed by defining the following functional
that includes the observational constraints (classical as well as asteroseismic) 
currently available for both components of 70~Ophiuchi
\begin{eqnarray}
\label{eq1}
\chi^2 \equiv \sum_{i=1}^{N_{\rm tot}} \left( \frac{C_i^{\mathrm{theo}}-
C_i^{\mathrm{obs}}}{E_i^{\mathrm{obs}}} \right)^2  \; ,
\end{eqnarray}
where $N_{\rm tot}$ is the total number of observational constraints included in the calibration.
The vectors $\mathbf{C}$ contains therefore all observables for both stars:  
\begin{eqnarray}
\nonumber
\mathbf{C} \equiv (M_{\rm A}, M_{\rm B}, L_{\rm A}, L_{\rm B}, 
T_{\mathrm{eff}}^{\rm A}, T_{\mathrm{eff}}^{\rm B}, 
[\mathrm{Fe/H}]_{\mathrm{s}}^{\rm A}, [\mathrm{Fe/H}]_{\mathrm{s}}^{\rm B}, \Delta \nu_0^{\rm A}) \; .   
\end{eqnarray} 
The vector $\mathbf{C}^{\mathrm{theo}}$ contains the theoretical values of these observables for the model to be tested, while 
the values of $\mathbf{C}^{\mathrm{obs}}$ are those
listed in Table~\ref{tab:constraints}. The vector $\mathbf{E^{\mathrm{obs}}}$ contains the errors on these observations, which are also given in Table~\ref{tab:constraints}.

\section{Results}
\label{res}

\begin{figure}[htb!]
 \resizebox{\hsize}{!}{\includegraphics{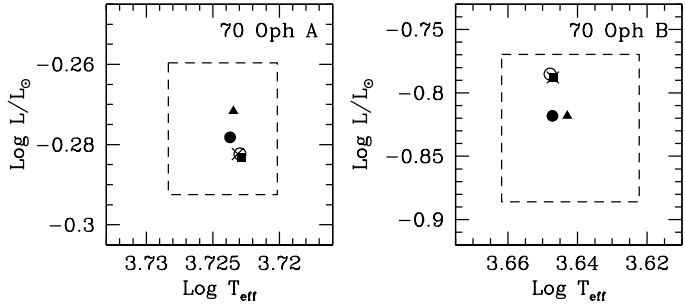}}
  \caption{Location in the HR diagram for the models of \object{70~Oph~A} and B. The dashed lines indicate the boxes delimited by the observed luminosities and effective temperatures (with their respective 1-sigma errors). The dot, the square and the triangle correspond to the M1a, M1b and M1c models, respectively. The open circle indicates the position of the M2 model, while the cross corresponds
to the M3 model. Note that the symbols corresponding to the M1b, M2 and M3 models are partly superimposed. In particular, the crosses corresponding to the M3 model are superimposed to the 
squares indicating the location of the M1b model.}
  \label{dhr}
\end{figure}

\begin{table*}
\caption{Models for \object{70~Oph~A} and B obtained with three different stellar evolution
codes. The M1 models are computed with the solar mixture of
\cite{gre93} and include atomic diffusion of helium and heavy elements. Model M2 is computed with the new solar mixture of
\cite{asp05} complemented with the neon abundance of \cite{cuh06} and includes atomic diffusion
of helium and heavy elements. Model M3 is computed with the solar mixture of \cite{gre93} but without atomic diffusion.
These models are computed with a solar-calibrated value of the mixing-length parameter.
The upper part of the table gives the non-asteroseismic observational constraints used for the calibration.
The middle part of the table gives the modeling parameters with their confidence limits, while the bottom part presents the global parameters of both stars.}
\begin{center}
\scriptsize
\label{tab:m1}
\begin{tabular}{c|cc|cc|cc|cc|cc}
\hline
\hline
 & \multicolumn{2}{c|}{Model M1a}  & \multicolumn{2}{c|}{Model M1b} & \multicolumn{2}{c|}{Model M1c} & \multicolumn{2}{c|}{Model M2}  & \multicolumn{2}{c}{Model M3}  \\
 & \multicolumn{2}{c|}{Geneva code}  & \multicolumn{2}{c|}{CLES code} & \multicolumn{2}{c|}{CESAM  code} & \multicolumn{2}{c|}{Solar mixture: Asp+Cunha}  & \multicolumn{2}{c}{No atomic diffusion} \\
 & \multicolumn{1}{c}{\object{70~Oph~A}} & \multicolumn{1}{c|}{\object{70~Oph~B}} & \multicolumn{1}{c}{\object{70~Oph~A}} & \multicolumn{1}{c|}{70~Oph B} & \multicolumn{1}{c}{\object{70~Oph~A}} & \multicolumn{1}{c|}{70~Oph B} & \multicolumn{1}{c}{\object{70~Oph~A}} & \multicolumn{1}{c|}{70~Oph B} & \multicolumn{1}{c}{\object{70~Oph~A}} & \multicolumn{1}{c}{70~Oph B} \\ 
\hline
$M/M_{\odot}$  & $0.89 \pm 0.02$ & $0.73 \pm 0.01$ & $0.89 \pm 0.02$ & $0.73 \pm 0.01$& $0.89 \pm 0.02$ & $0.73 \pm 0.01$& $0.89 \pm 0.02$ & $0.73 \pm 0.01$& $0.89 \pm 0.02$ & $0.73 \pm 0.01$   \\
$L/L_{\odot}$ & $ 0.53 \pm 0.02$ & $0.15 \pm 0.02$ & $ 0.53 \pm 0.02$ & $0.15 \pm 0.02$& $ 0.53 \pm 0.02$ & $0.15 \pm 0.02$& $ 0.53 \pm 0.02$ & $0.15 \pm 0.02$& $ 0.53 \pm 0.02$ & $0.15 \pm 0.02$  \\
$T_{\mathrm{eff}}$ (K)& $5300 \pm 50$ & $4390 \pm 200$ & $5300 \pm 50$ & $4390 \pm 200$& $5300 \pm 50$ & $4390 \pm 200$& $5300 \pm 50$ & $4390 \pm 200$& $5300 \pm 50$ & $4390 \pm 200$  \\
$[$Fe/H$]$ &  \multicolumn{2}{c|}{$0.04 \pm 0.05$}  &  \multicolumn{2}{c|}{$0.04 \pm 0.05$}&  \multicolumn{2}{c|}{$0.04 \pm 0.05$}&  \multicolumn{2}{c|}{$0.04 \pm 0.05$}&  \multicolumn{2}{c}{$0.04 \pm 0.05$}\\ 
\hline
$M/M_{\odot}$  & $0.91 \pm 0.02$ & $0.72 \pm 0.01$ & $0.90 \pm 0.02$ & $0.73 \pm 0.01$ & $0.91 \pm 0.02$ & $0.72 \pm 0.01$ & $0.90 \pm 0.02$ & $0.73 \pm 0.01$ & $0.90 \pm 0.02$ & $0.73 \pm 0.01$ \\
$t$ (Gyr) & \multicolumn{2}{c|}{$6.2 \pm 1.0$} & \multicolumn{2}{c|}{$6.5 \pm 1.0$} & \multicolumn{2}{c|}{$6.2 \pm 1.0$} & \multicolumn{2}{c|}{$6.3 \pm 1.0$} & \multicolumn{2}{c}{$7.2 \pm 1.2$}   \\
$Y_{\mathrm{i}}$ & \multicolumn{2}{c|}{$0.266 \pm 0.015$} & \multicolumn{2}{c|}{$0.266 \pm 0.018$} & \multicolumn{2}{c|}{$0.272 \pm 0.015$} & \multicolumn{2}{c|}{$0.255 \pm 0.018$} & \multicolumn{2}{c}{$0.254 \pm 0.018$}  \\
$(Z/X)_{\mathrm{i}}$ & \multicolumn{2}{c|}{$0.0300 \pm 0.0025$} & \multicolumn{2}{c|}{$0.0300 \pm 0.0025$} & \multicolumn{2}{c|}{$0.0300 \pm 0.0025$} & \multicolumn{2}{c|}{$0.0220 \pm 0.0023$} & \multicolumn{2}{c}{$0.0266 \pm 0.0025$} \\
\hline
$L/L_{\odot}$ & 0.527 & 0.152 & 0.521 & 0.163 & 0.535 & 0.152 & 0.522 & 0.164 & 0.522 & 0.163  \\
$T_{\mathrm{eff}}$ (K)& $5293$ & $4438$ & $5283$ & $4434$ & $5290$ & $4395$ & $5284$ & $4445$ & $5286$ & $4437$ \\
$R/R_{\odot}$ & $0.865$ & $0.661$ & $0.863$ & $0.685$ & $0.872$ & $0.674$ & $0.864$ & $0.684$ & $0.863$ & $0.684$ \\
$Y_{\mathrm{s}}$ & 0.238 & 0.248 & 0.238 & 0.247 & 0.244 & 0.253 & 0.228 & 0.237 & 0.254 & 0.254  \\
$(Z/X)_{\mathrm{s}}$ & 0.0269 & 0.0280 & 0.0267 & 0.0279 & 0.0266 & 0.0277 & 0.0192 & 0.0201 & 0.0266 & 0.0266  \\
$[$Fe/H$]_{\mathrm{s}}$ &  $0.04$ & $0.06$ &  $0.04$ & $0.06$ &  $0.04$ & $0.05$ &  $0.04$ & $0.06$ &  $0.04$ & $0.04$ \\
\hline
\end{tabular}
\end{center}
\end{table*}

Using the observational constraints listed in Sect.~\ref{obs},
we perform the aforementioned $\chi^2$ minimization.
In a first time, we fix the mixing-length parameter to its
solar calibrated value: $\alpha_{\rm A}=\alpha_{\rm B}=\alpha_{\odot}$.
The solar mixture of \cite{gre93} is used and atomic diffusion of
helium and heavy elements is included in the computation.

The calibration is first performed by computing a dense grid of stellar
models with the Geneva stellar evolution code. 
In this way, we find the following solution
$t=6.2 \pm 1.0$\,Gyr, $M_{\rm A}=0.91 \pm 0.02$\,M$_{\odot}$, $M_{\rm B}=0.72 \pm 0.01$\,M$_{\odot}$, 
$Y_{\mathrm{i}}=0.266 \pm 0.015$ and $(Z/X)_{\mathrm{i}}=0.0300 \pm 0.0025$. 
The position in the HR diagram
of this model of \object{70~Oph~A} and B (denoted model M1a in the following) is given in Fig.~\ref{dhr}. 
The characteristics of this model
are reported in Table~\ref{tab:m1}. 
The confidence limits of each modeling parameter given in Table~\ref{tab:m1} are estimated as the maximum/minimum values that fit the observational constraints when the other calibration parameters are fixed to their medium value.

In order to compare results obtained by using a different stellar evolution code, 
the same calibration is done by using the CLES code and the Levenberg-Marquardt method.
This calibration (called model M1b) leads to
the following solution: $t=6.5 \pm 1.0$\,Gyr, 
$M_{\rm A}=0.90 \pm 0.02$\,M$_{\odot}$, $M_{\rm B}=0.73 \pm 0.01$\,M$_{\odot}$, 
$Y_{\mathrm{i}}=0.266 \pm 0.018$ and $(Z/X)_{\mathrm{i}}=0.0300 \pm 0.0025$.
The characteristics of this solution are given in Table~\ref{tab:m1} and the location
of both components in the HR diagram is shown in Fig.~\ref{dhr}.
A third comparison is finally performed by using the CESAM evolution code.
The following solution (denoted model M1c) is then found:
$t=6.2 \pm 1.0$\,Gyr, 
$M_{\rm A}=0.91 \pm 0.02$\,M$_{\odot}$, $M_{\rm B}=0.72 \pm 0.01$\,M$_{\odot}$, 
$Y_{\mathrm{i}}=0.272 \pm 0.015$ and $(Z/X)_{\mathrm{i}}=0.0300 \pm 0.0025$.
The global parameters of \object{70~Oph~A}B determined with this solution are also given
in the bottom part of Table~\ref{tab:m1}, while Fig.~\ref{dhr} shows the location
of both components in the HR diagram.

\begin{figure}[htb!]
 \resizebox{\hsize}{!}{\includegraphics{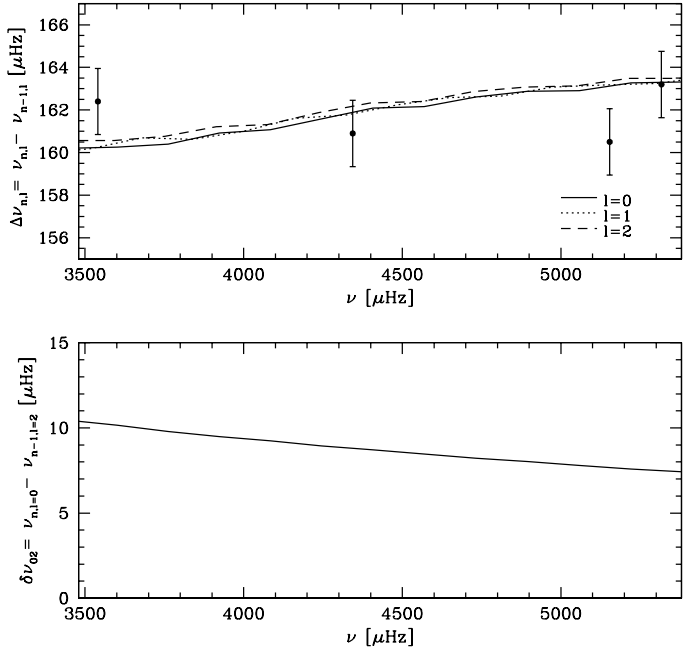}}
  \caption{Large and small separations versus frequency for the M1b model of \object{70~Oph~A}. The dots indicate the observed values of the large separation with an uncertainty on
  individual frequencies estimated to half the time resolution (1.1\,$\mu$Hz).}
  \label{gdpt_CLES}
\end{figure}

From these results, we first conclude that a consistent model of the 70~Ophiuchi system that
correctly reproduces all observational constraints now available for \object{70~Oph~A} and
B can be determined. Indeed, Fig.~\ref{dhr} and Table~\ref{tab:m1} 
show that the three solutions are in very good agreement with 
all classical observables included in the calibration. 
Moreover, the M1 models correctly reproduce the observed value 
of the mean large separation of \object{70~Oph~A}. 
Note that the global asteroseismic features of the three M1 models of \object{70~Oph~A} are
very similar; they all exhibit a mean large separation of about $161.8$\,$\mu$Hz
with a mean small separation between $\ell=2$ and $\ell=0$ of about $9.5$\,$\mu$Hz.
As an illustration, the variation in the large and small separation $\delta \nu_{02}$ 
with frequency for the M1b model of \object{70~Oph~A} is shown in Fig.~\ref{gdpt_CLES}.
From Table~\ref{tab:m1} we also conclude that the three different stellar evolution
codes give very similar results. We only note a slightly higher value of the initial helium
mass fraction determined with the CESAM evolution code, which is directly related to the
higher value of this parameter found for the solar calibration (see Sect.~\ref{phys}). 
These results are comforting since the input physics included
in these three codes is similar (although not striclty identical, see Sect.~\ref{phys})
and must therefore lead to a coherent determination of the modeling parameters of
the 70~Ophiuchi system. It is also worthwhile to recall here that, in addition to the
stellar evolution codes, the calibration methods were compared. Indeed, the 
M1a model has been obtained by computing a dense grid of stellar models, while
the solution M1b has been determined by using an optimization algorithm. We thus find that
these methods lead to the same determination of the global parameters of \object{70~Oph~A}B.
These results are in good agreement with previous computations done for the 
calibration of the $\alpha$~Centauri system by using the Geneva and the CLES evolution code \citep[see][]{egg04_acen,mig05}. We also note that the values of the parameters that we found for the 70~Ophiuchi system are globally in good agreement with the results of previous studies by \cite{fer98} and
\cite{cas07} with smaller error bars resulting from the numerous observational constraints now available for
this system.

Thus an age of about 6\,Gyr with a typical uncertainty of 1\,Gyr is determined for the 70~Ophiuchi system 
independently of the evolution code used for the computation.
Concerning the initial chemical
composition, we find an initial ratio between the mass fraction of heavy elements and hydrogen 
$(Z/X)_{\mathrm{i}} = 0.0300 \pm 0.0025$, which is slightly higher than for a solar model.
This value is directly constrained by the observed surface metallicity
$[$Fe/H$]_{\mathrm{s}}=0.04 \pm 0.05$. Contrary to the initial ratio between the mass fraction of heavy elements and hydrogen, a slightly lower value of the initial helium abundance is found
for \object{70~Oph~A}B than for a solar model. We however note that the error on this parameter is large.
Moreover, we recall here that all these results have been obtained by fixing
the mixing-length parameter of both components of the 70~Ophiuchi system
to the solar calibrated value. We are now interested in investigating
which solutions can be found by relaxing this constraint.

\subsection{Models with a free mixing-length parameter}
\label{freemlt}

\begin{figure}[htb!]
 \resizebox{\hsize}{!}{\includegraphics{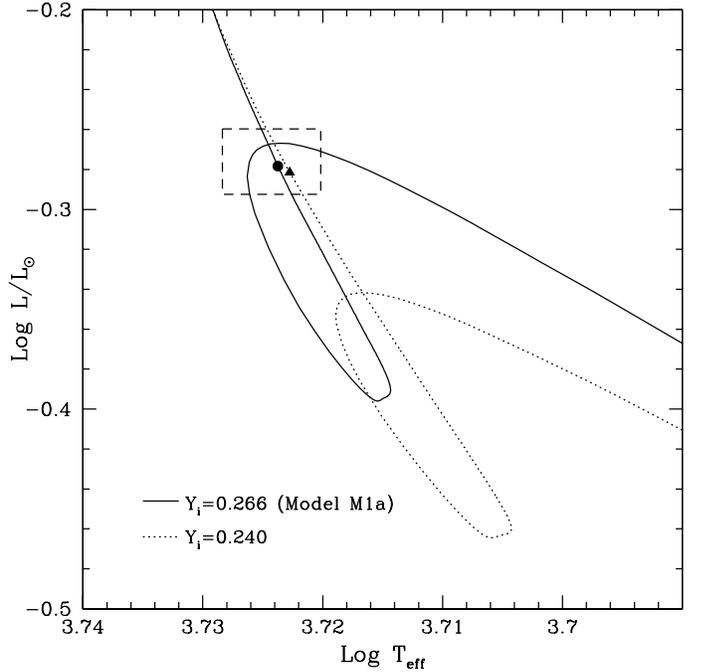}}
  \caption{Evolutionary tracks in the HR diagram for two models of \object{70~Oph~A} with different
values of the initial helium mass fraction $Y_{\mathrm{i}}$ and convection parameter $\alpha$.
The model M1a computed with $Y_{\mathrm{i}}=0.266$ has a solar mixing-length parameter, while
the solution with a lower initial helium mass fraction of $Y_{\mathrm{i}}=0.240$ exhibits a higher value of this parameter ($\alpha=1.25 \alpha_{\odot}$).
This shows that a decrease in the initial helium mass fraction can be compensated
by an increase of the mixing-length parameter in order to reach the same location in the HR diagram.}
  \label{dhr_yi}
\end{figure}

The calibration of the 70~Ophiuchi system is then done by relaxing the constraint of
a solar mixing-length parameter. We however still assume that the mixing-length parameter of \object{70~Oph~A} is equal to the one of the B component. We then find that models with different values of the initial helium abundance are able to correctly reproduce all
observational constraints now available for the 70~Ophiuchi system. 
This is due to the fact that a decrease (increase) of the 
initial helium abundance $Y_{\mathrm{i}}$ can be compensated by an increase (decrease) 
of the mixing-length parameter $\alpha$ to match the observed position of \object{70~Oph~A} in the HR diagram 
as found previously for the calibration of the star Procyon~A
\citep[see][]{egg05_pro}. As an illustration, Fig.~\ref{dhr_yi}
shows the evolutionary track of \object{70~Oph~A} for a model computed with an
initial helium abundance $Y_{\mathrm{i}}=0.240$ and a mixing-length parameter
$\alpha=1.25 \alpha_{\odot}$. This model shares approximately the same location
in the HR diagram as the M1a model but exhibits a much higher age of about 10.5\,Gyr  
due to its very low initial helium abundance. 
We thus obtain a series of models with approximately the same non-asteroseismic features as those of the M1 models computed with a solar calibrated value of the mixing-length parameter.  
Moreover, the mean large separation of these models (which is the only asteroseismic quantity included in the $\chi^2$ functional) is also very close to the value of the M1a model, since it mainly depends on the square root of the star's mean density.

\begin{figure}[htb!]
 \resizebox{\hsize}{!}{\includegraphics{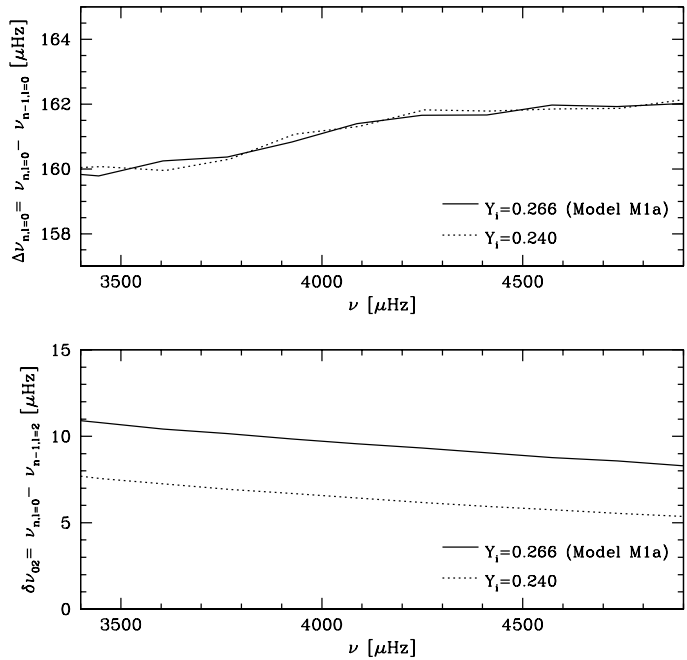}}
  \caption{Large and small separations corresponding to the models of 
\object{70~Oph~A} whose location in the HR diagram is shown in Fig.~\ref{dhr_yi}.
The solution with $Y_{\mathrm{i}}=0.266$ has a solar mixing-length parameter, while
the model with $Y_{\mathrm{i}}=0.240$ has been computed with $\alpha=1.25 \alpha_{\odot}$.}
  \label{gdpt_yi}
\end{figure}

\begin{figure}[htb!]
 \resizebox{\hsize}{!}{\includegraphics{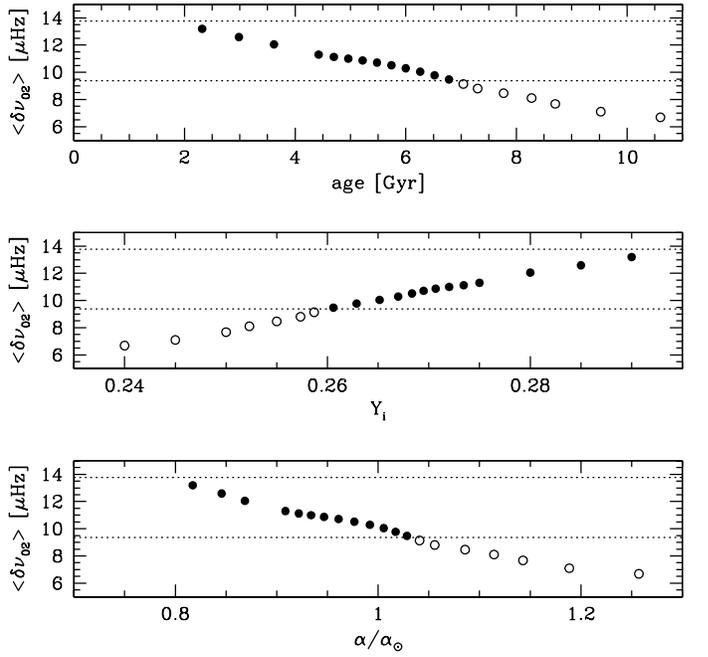}}
  \caption{Mean small separation $\delta \nu_{02}$ of the component A for
           different models of 70 Ophiuchi. Dots and open symbols correspond to solutions
           with an age lower and higher than 7\,Gyr, respectively.
           Dotted lines indicate the allowed frequency interval 
           for the mean small separation suggested by current asteroseismic observations.}
  \label{dnu02_vs_t}
\end{figure}

The observed location of \object{70~Oph~B} in the HR diagram only provides
weak constraints to discriminate between these models because of the large error on its
effective temperature and because the model of \object{70~Oph~B} spend most of its time
within the observational box in the HR diagram during its main-sequence evolution
due to its low mass of 0.73\,M$_{\odot}$. Only solutions with an initial helium
abundance $Y_{\mathrm{i}}$ larger than about 0.290 (corresponding to an age lower
than about 2\,Gyr) lead to models of \object{70~Oph~B} with a too high luminosity to reproduce
the classical observations. 
This explains why there is a series of models of \object{70~Oph~A}B with very different ages
and initial helium abundances that well reproduce the global stellar parameters considered in the calibration. We however note that these models exhibit different asteroseismic features 
although they are characterized by the same value of the mean large separation.
In particular, these models have different values of the mean small separation 
$\delta \nu_{02}$ since their ages differ. This is shown in Fig.~\ref{gdpt_yi}
where the large and small separations of \object{70~Oph~A} for the M1a model 
and for the model computed with $Y_{\mathrm{i}}=0.240$ are compared. We see that both
models are characterized by the same values of the large separations, while the mean small
separation $\delta \nu_{02}$ of the M1a model is significantly larger (3\,$\mu$Hz) 
than the one of the model with a lower initial helium abundance. This illustrates the
importance of having a precise observed value of the mean small separation of \object{70~Oph~A}  
in order to obtain an independent determination of the value of the age, 
the mixing-length parameter and the initial helium abundance of the 70~Ophiuchi system.
This is illustrated in more detail in Fig.~\ref{dnu02_vs_t} where the value of the
mean small separations of \object{70~Oph~A} for models of \object{70~Oph~A}B computed for different 
initial helium abundances and mixing-length parameters is shown. As expected, 
the mean small separation is found to significantly decrease with the age. The increase of the
age of the system is directly correlated to the decrease of the initial helium abundance, 
which is in turn directly related to the increase of the value of mixing-length parameter 
needed to reproduce the observed location of \object{70~Oph~A} in the HR diagram and the observed
mean large separation. We thus conclude that the observation of the mean small
separation of \object{70~Oph~A} will lift the degeneracy between the value of the mixing-length parameter 
and the initial helium abundance by adding a strong observational constraint
on the age of the system and will therefore enable an independent and precise
determination of the mixing-length parameter and the initial helium abundance 
of the 70~Ophiuchi system. 

The mean small separation of \object{70~Oph~A} has unfortunately not been observed. 
The mono-site nature of current asteroseismic observations 
coupled to the low resolution of the time series do not enable such a
measurement. As discussed in Sect.~\ref{asc}, these observations nevertheless suggest that the value
of the mean small separation of \object{70~Oph~A} is included in a frequency interval centered
around the daily alias of 11.57\,$\mu$Hz introduced by the mono-site observations 
and delimited by the frequency resolution of the time-series (2.2\,$\mu$Hz).
As shown in Fig.~\ref{dnu02_vs_t}, 
this constraint on the small separation rules out models of
\object{70~Oph~A}B with an age higher than about 7\,Gyr, an initial helium abundance
$Y_{\mathrm{i}}$ lower than about 0.260 and a mixing-length parameter $\alpha$ larger
than about 1.05$\alpha_{\odot}$. We thus conclude that current asteroseismic observations
suggest that the age of the 70~Ophiuchi system lies between 2 and 7\,Gyr with a mixing-length
parameter between about 0.8 and 1.05$\alpha_{\odot}$ and an initial initial helium abundance
$Y_{\mathrm{i}}$ between 0.260 and 0.290 when the constraint of
a solar mixing-length parameter is relaxed.

\subsection{Models with the new solar mixture}

In order to test the sensitivity of the global parameters 
determined for 70 Oph AB on the adopted solar mixture, we
decided to redo the whole calibration using the solar mixture proposed by
\cite{asp05} complemented with the neon abundance of \cite{cuh06}.
This results in a lower solar value of the ratio between the mass fraction 
of heavy elements and hydrogen $(Z/X)_{\odot}=0.0178$. We recall here
that solar models evolved with this new abundance mixture give worse
agreement with helioseismic constraints \citep[see e.g.][]{guz06}.
For these computations, we use OPAL opacity tables calculated with the 
corresponding solar mixture and the mixing-length parameter
is fixed to its solar calibrated value. We then find the
following solution: $t=6.3 \pm 1.0$\,Gyr, $M_{\rm A}=0.90 \pm 0.02$\,M$_{\odot}$, $M_{\rm B}=0.73 \pm 0.01$\,M$_{\odot}$, 
$Y_{\mathrm{i}}=0.255 \pm 0.018$ and $(Z/X)_{\mathrm{i}}=0.0220 \pm 0.0023$. 
The position in the HR diagram
of this model of \object{70~Oph~A} and B (denoted model M2 in the following) is given in Fig.~\ref{dhr} and 
its global features are reported in Table~\ref{tab:m1}. 
We note that this solution calibrated with the new solar abundances presents a
different initial chemical composition, but no other significant
deviation from the global parameters of the M1 models computed with the solar mixture of
\cite{gre93}. Indeed, the M1 and M2 models share the same age and the same locations in the HR diagram.
A lower value of the initial helium abundance is then found for the M2 model in order to compensate
for the decrease in the initial ratio between the mass fraction 
of heavy elements and hydrogen resulting from the lower solar value $(Z/X)_{\odot}$.
As it happens for the Sun and for $\alpha$ Centauri A \citep[see][]{mig05}, the models of
\object{70~Oph~A} computed with the solar mixture of \cite{gre93} exhibit a
deeper convective zone compared with the M2 model, but the uncertainty in the observational
constraints do not enable to reveal this difference that can be masked by other choice
of parameters.

\subsection{Models without atomic diffusion}

The effects of atomic diffusion on the calibration of the 70 Ophiuchi system are also
studied by computing models without diffusion. For these computations, the solar mixture
of \cite{gre93} is used and the mixing-length parameter
is fixed to its solar calibrated value. Thus the same calibration is made as for the M1 models
except for the inclusion of atomic diffusion. We then obtain the solution (hereafter refered as M3)
$t=7.2 \pm 1.2$\,Gyr, $M_{\rm A}=0.90 \pm 0.02$\,M$_{\odot}$, $M_{\rm B}=0.73 \pm 0.01$\,M$_{\odot}$, 
$Y_{\mathrm{i}}=0.254 \pm 0.018$ and $(Z/X)_{\mathrm{i}}=0.0266 \pm 0.0025$. 
The characteristics of this solution are given in Table~\ref{tab:m1} while the location
of both components in the HR diagram is shown in Fig.~\ref{dhr}.
By comparing model M3 with the M1 models, we see that the initial chemical composition
differs. Indeed, the M1 models are characterized by a higher initial 
ratio between the mass fraction of heavy elements and hydrogen than the M3 solution
because of the inclusion of atomic diffusion. Moreover the initial helium abundance of
the model without diffusion is lower and this results in an increase of the age of about 1\,Gyr.  

\begin{figure}[htb!]
 \resizebox{\hsize}{!}{\includegraphics{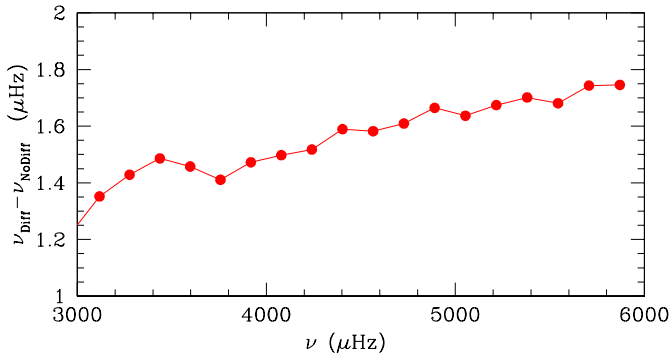}}
  \caption{Differences of frequencies between the model of \object{70~Oph~A} computed with the
inclusion of atomic diffusion (solution M1b) and the model of \object{70~Oph~A} computed without
atomic diffusion (solution M3).}
  \label{dfreq}
\end{figure}

Figure~\ref{dfreq} shows the effects of atomic diffusion on the frequencies of \object{70~Oph~A}.
The differences between the frequencies of the M2 and the M1b models of \object{70~Oph~A}
(defined as $\nu_{n0}({\rm M1b})-\nu_{n0}({\rm M3})$) show that the slightly lower
value of the surface abundance in helium of the model computed without atomic
diffusion (difference of about 4\,\%) does not significantly change the value of the frequencies.
For a low-mass star like \object{70~Oph~A}, the effects of atomic diffusion are indeed small
because of the large mass contained in its convective zone. 
We thus conclude that the change in the frequencies introduced by
the different value of the surface helium abundance that implies a different opacity and, therefore, a different location of the boundary of the convective zone is too small to be clearly revealed.

\section{Conclusion}

Our calibration of the 70~Ophiuchi AB shows that a solution correctly reproducing
all asteroseismic and non-asteroseismic observational constraints now available for both stars
can be determined. An age of $6.2 \pm 1.0$\,Gyr is found with an initial
helium mass fraction $Y_{\mathrm{i}}=0.266 \pm 0.015$ and an initial metallicity 
$(Z/X)_{\mathrm{i}}=0.0300 \pm 0.0025$ when atomic diffusion is included and
a solar value of the mixing-length parameter is assumed. 
Concerning the mixing-length parameter,
we found that current asteroseismic measurements do not allow a firm 
independent determination of this value that will only be possible with a precise
measurement of the mean small separation.  
The effects of atomic diffusion and of the choice of the adopted solar mixture on the results of the calibration were also studied.
We then found that a change of the solar mixture results of course in a
different initial chemical composition but does not change significantly
the other global parameters. Models computed without atomic diffusion
also have a different initial chemical composition than models including
diffusion but they also exhibit a slighlty higher age (difference of about
1\,Gyr). 

We finally mention that we took the opportunity of the calibration of the
70~Ophiuchi system to compare and test the theoretical
tools used for the modeling of stars for which p-modes frequencies are 
detected. Thus, the same analysis was performed with three different stellar
evolution codes including a coherent input physics: the Geneva code, the CLES code
and the CESAM code. Two different calibration methods were also used: 
the computation of a dense grid of stellar models and a minimization algorithm.  
We found that the different evolution codes and calibration methods lead
to perfectly similar and coherent solutions. This result is comforting in the perspective
of the theoretical interpretation of asteroseimic data that are expected from the
CoRoT and the Kepler space missions and from future ground-based campaigns.

\begin{acknowledgements}
We thank J. Christensen-Dalsgaard for providing us with the Aarhus adiabatic pulsation code. 
This work has made use of the ORBIT code developed by T. Forveille.
A.M. acknowledges financial support from the Prodex-ESA Contract Prodex 8 COROT (C90199).
Part of this work was supported by the Swiss National Science Foundation.
\end{acknowledgements}

\bibliographystyle{aa} 
\bibliography{biblio} 

\end{document}